\documentclass[a4paper,notitlepage,twocolumn,footnoteinbib,longbibliography,superscriptaddress,10pt]{revtex4-1}

\usepackage{amsmath,amssymb,amsthm,mathtools,color,listings,graphicx}
\usepackage[colorlinks=true,linkcolor=blue,citecolor=blue]{hyperref}
\usepackage{braket}

\DeclareMathAlphabet{\mathbbb}{U}{bbold}{m}{n}

\begin{document}

\title{Ising selector machine by Kerr parametric oscillators}
\author{Jacopo Tosca}
\affiliation{Universit\'e Paris Cit\'e, CNRS, Mat\'eriaux et Ph\'enom\'enes Quantiques, 75013 Paris, France}
\author{Cristiano Ciuti}
\affiliation{Universit\'e Paris Cit\'e, CNRS, Mat\'eriaux et Ph\'enom\'enes Quantiques, 75013 Paris, France}
\author{Claudio Conti}
\affiliation{Physics Department, Sapienza University of Rome, 00185 Rome, Italy}
\affiliation{Centro Ricerche Enrico Fermi (CREF), Via Panisperna 89a, 00184 Rome, Italy}
\author{Marcello Calvanese Strinati}
\email{marcello.calvanesestrinati@gmail.com}
\affiliation{Centro Ricerche Enrico Fermi (CREF), Via Panisperna 89a, 00184 Rome, Italy}
\date{\today}

\begin{abstract}
Ising machines are physical platforms designed to minimize the energy of classical Ising Hamiltonians, yet accessing specific excited states remains an open challenge of both fundamental and practical relevance. In this letter we show that a network of Kerr parametric oscillators (KPOs) naturally implements an Ising selector machine. By tuning the frequency detuning between the parametric pump and the oscillator resonances, the system can be steered to converge close to the ground state, the highest-energy configuration, or targeted intermediate excited states. Beyond mean field, numerical simulations based on the truncated Wigner approximation demonstrate that noise insertion preserves the energetic structure of the landscape. The targeted state emerges with an exponentially enhanced probability over the rest of the Ising spectrum. Our results establish the pump-cavity detuning as a control knob for navigating the full Ising energy landscape, opening a route to applications in Boltzmann sampling, hardness characterization, and spectral analysis of combinatorial problems.
\end{abstract}

\maketitle

\textit{Introduction} -  
A wide variety of combinatorial optimization problems can be formulated as the minimization of a classical Ising energy~\cite{Lucas2014,Barahona1982}. 
Since this task belongs to the NP-hard complexity class, with the computational cost growing exponentially with the system size, 
a number of physical platforms, known as Ising machines, have been developed to efficiently find low-energy spin configurations. 
Prominent realizations include networks of optical parametric oscillators (OPOs)~\cite{yamamoto2020isingmachine,Inagaki2016,Honjo2021}, superconducting circuits, electronic oscillators, and spatial photonic systems (see Ref.~\cite{Mohseni2022} for a review). 
To date, these architectures have been designed for a single goal: Reaching the ground state of the encoded Ising Hamiltonian.
 
However, access to the full energy landscape of the Ising model, and not just its minimum, is of considerable fundamental and practical interest.
Determining the energy of the $n$th excited state is itself an NP-hard problem, as it generally requires knowledge of the entire configuration space. 
Yet excited-state information is essential in a number of contexts: 
Evaluating spectral gaps and the density of low-lying states provides direct insight into the computational hardness of an optimization instance and governs the performance of annealing algorithms~\cite{Barahona1982}. 
Sampling spin configurations at controlled energy is central to Boltzmann machine learning and maximum-entropy inference~\cite{Goto2018, vandersande2022}, and in statistical physics, the structure of the excitation spectrum determines thermodynamic properties and relaxation pathways.
A physical device capable of \emph{selecting} a target Ising state at a prescribed energy level, rather than merely minimizing the energy, would therefore constitute a qualitatively new tool, going beyond the current paradigm of Ising machines as pure optimizers.
 
In this work, we introduce and analyze such a state-selection mechanism and show that it emerges naturally in networks of Kerr parametric oscillators (KPOs). 
KPOs have recently attracted considerable attention as a platform for encoding and solving Ising problems~\cite{goto2019kpoandopo,Goto2016,Nigg2017,Kanao2021}, 
with experimental implementations in superconducting circuits~\cite{Yamaji2023,lvarez2024} and photonic systems~\cite{Okawachi2020}. 
Prior studies have explored these systems in minimal configurations, studying mean-field attractors and demonstrating that stochastic dynamics facilitate the emergence of the ground state~\cite{lvarez2024}.

We demonstrate that the frequency detuning between the parametric pump and the oscillator resonances acts as a selector that steers the system toward different eigenstates of the coupling matrix, and hence toward different Ising energy levels. 
For different detuning, the network converges close to the ground state, the highest-energy configuration, or specific intermediate states. 
We establish this result analytically at the mean-field level and confirm it numerically using the truncated Wigner approximation (TWA)~\cite{polkovnikov2010}, which captures quantum fluctuations beyond mean field. 
Crucially, the presence of noise preserves the energetic macro-structure of the landscape: The targeted state, whether it is the ground state or an excited one, emerges with an exponentially enhanced probability relative to the rest of the Ising spectrum~\cite{Tosca2025}.

\textit{Model} -
A KPO consists of a parametric oscillator with proper frequency $\omega_0$ and Kerr nonlinearity $U>0$, which is subject to an effective two-photon pump with frequency $\omega_{\rm p}$ and amplitude $G>0$.  In the reference frame rotating at half the pump frequency, and in the rotating wave approximation, the system Hamiltonian is given by ($\hbar=1$)~\cite{goto2019kpoandopo}
\begin{equation}
\hat H_0(\hat a)=-\Delta\,\hat a^\dag\hat a+\frac{U}{2}\,\hat a^\dag\hat a^\dag\hat a\hat a+\frac{G}{2}\left(\hat a^\dag\hat a^\dag+\hat a\hat a\right) \,\, .
\label{eq:nonlinearmodelisingmachine01}
\end{equation}
Here, $\Delta=\omega_0-\omega_{\rm p}/2$ is the frequency detuning from the reference frequency $\omega_{\rm p}/2$. The operators $\hat a$ ($\hat a^\dag$) are the bosonic annihilation (creation) operator, obeying the canonical commutation relations $[\hat a,\hat a^\dag]=1$ and $[\hat a,\hat a]=~0$. 
The evolution of the system density matrix $\hat\rho$, in the presence of weak coupling to
a Markovian environment,  is given by the Lindblad master equation
\begin{equation}
\frac{d\hat\rho}{dt}=-i\left[\hat H_0(\hat a),\hat\rho(t)\right]+{\mathcal{D}}\hat\rho(t) \,\, ,
\label{eq:nonlinearmodelisingmachine01bis001}
\end{equation}
where the superoperator $\mathcal D$ encodes one-photon dissipation at rate $\gamma$ via the jump operator $\hat \Gamma = \sqrt{\gamma}\,\hat a$ as 
\begin{equation}
    \mathcal{ D} {\hat \rho(t)}=
 {\hat \Gamma} {\hat \rho(t)} \hat \Gamma_{}^{\dagger}-\frac{1}{2}
\left\{
\hat \Gamma_{}^{\dagger} \hat \Gamma_{}, 
{\hat \rho(t)}
\right\} \,\, .
\label{eq:nonlinearmodelisingmachine01bis003}
\end{equation}
The reason why a single KPO may simulate an Ising spin can be understood 
by considering the mean-field dynamics of the bosonic mode, namely $A(t) = {\rm Tr}[{\hat{\rho}(t) \hat{a}}]$, given by the equation:  
\begin{equation}
\frac{d A}{dt}=\frac{1}{i}\left(-\Delta  A  + G  A^* +U{|A|}^2 A \right)-\frac{\gamma}{2}A \,\, .
\label{eq:nonlinearmodelisingmachine01bis004}
\end{equation}
Separating $A$ into its real and imaginary parts as $A = X + iY$, and introducing the polar coordinates $X = R\cos(\varphi)$ and $Y = R\sin(\varphi)$, the steady state with $R > 0$ is given by
\begin{subequations}
\begin{align}
R&=\sqrt{\frac{1}{U}\left(\Delta+\sqrt{G^2-\frac{\gamma^2}{4}}\,\right)} ,\\
\tan(\varphi)&=\frac{\gamma/2}{\sqrt{G^2-\gamma^2/4}-G} \,\, ,
\end{align}
\label{eq:nonlinearmodelisingmachine01bis006}
\end{subequations}
for $G$ above the oscillation threshold value $G_{\rm th}=\sqrt{\gamma^2/4+\Delta^2}$. By contrast, for $G < G_{\rm th}$, the system is below threshold and the stable solution is trivial, with $R = 0$ (the origin of the complex quadrature space). Equation~\eqref{eq:nonlinearmodelisingmachine01bis006} shows that the phase of the KPO, $\varphi$, admits two values: $\varphi$ and $\varphi \pm \pi$, where $\varphi \in (\pi/2,\pi)$ depends on $\gamma$ and $G$. These two phase values correspond to the ``up'' and ``down'' states of an isolated Ising spin.

\begin{figure}[t]
\centering
\includegraphics[width=8.6cm]{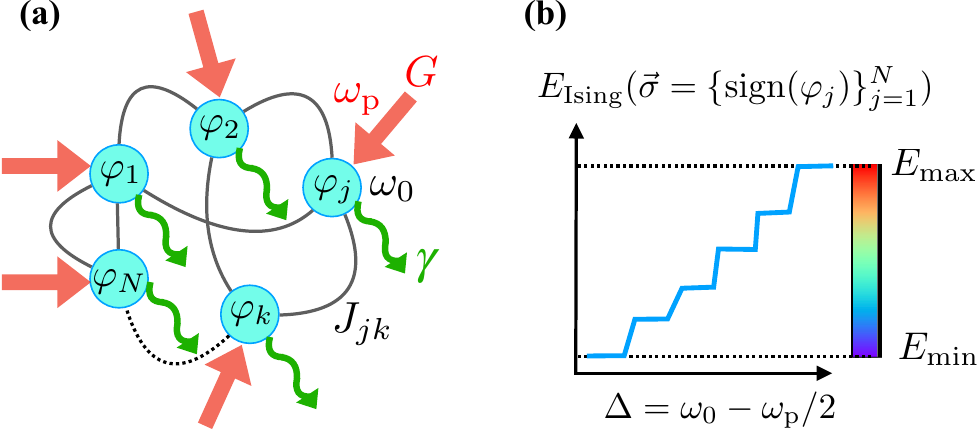}
\caption{\textbf{(a)} Schematic of the KPO network described by Eq.~\eqref{eq:nonlinearmodelisingmachine01bis008}. Each KPO (blue circle) has a resonance frequency $\omega_0$ and is driven at frequency $\omega_{\rm p}$ with amplitude $G$ (red arrows). Intrinsic single-photon dissipation occurs at rate $\gamma$ (green arrows). Different KPOs $j$ and $k$ are coupled via a real symmetric matrix $\mathbf{J}$, whose element $J_{jk}$ sets the interaction strength (grey lines). \textbf{(b)} Operating principle of the KPO network as an Ising selector machine. The phases $\varphi_j$ of the KPOs define an Ising spin configuration $\sigma_j={\rm sign}(\varphi_j)$. The corresponding Ising energy [Eq.~\eqref{eq:singlekerrparametrioscillatorising1}] near the oscillation threshold increases with the detuning $\Delta=\omega_0-\omega_{\rm p}/2$. This enables $\Delta$ to be used as a control parameter to select the Ising state toward which the KPO dynamics converges.}
\label{fig:networkandisingstateselection01}
\end{figure}

We now consider a system of $N$ KPOs coupled via the hopping Hamiltonian $\hat H_1=-(1/2)\sum_{j,k=1}^{N}J_{jk}\hat a^\dag_j\hat a_k$, where $\hat a_j$ is the bosonic annihilation operator for the $j$th mode and $\mathbf{J}$ is a symmetric coupling matrix whose element $J_{jk}$ gives the coupling strength between KPO $j$ and KPO $k$ (Fig.~\ref{fig:networkandisingstateselection01}\textbf{a}). In this case, Eq.~\eqref{eq:nonlinearmodelisingmachine01bis001} takes the form
\begin{equation}
\begin{aligned}
\frac{d\hat\rho}{dt}=-i\left[\sum_{j=1}^{N}\hat H_0(\hat a_j)+\hat H_1,\hat\rho\right]+\sum_{j=1}^{N}\hat{\mathcal{D}}_j\hat\rho(t) \,\, ,
\label{eq:nonlinearmodelisingmachine01bis007}
\end{aligned}
\end{equation}
where $\hat{\mathcal{D}}_j\hat\rho(t)$ is defined as in Eq.~\eqref{eq:nonlinearmodelisingmachine01bis003}, with jump operator $\hat \Gamma_j=\sqrt{\gamma}\,\hat a_j$. 
In this case, for each mode labeled by the index $j$, we introduce the mean-field amplitude $A_j=\mathrm{Tr}[\hat \rho\, \hat{a}_j]=X_j+i\,Y_j$. At the mean-field level, the dynamics is then governed by
\begin{equation}
\frac{d A_j}{dt}\!=\!\frac{1}{i}\left(\!-\Delta A_j\!+\!G A_j^*\!+\!U|A_j|^2A_j\!-\!\frac{1}{2}\sum_{k=1}^{N}J_{jk} A_k\!\right)\!-\frac{\gamma}{2}\!A_j. 
\label{eq:nonlinearmodelisingmachine01bis007MF}
\end{equation}
The dynamics of the real and imaginary parts $X_j$ and $Y_j$ can then be expressed in compact block form by introducing $\vec{X}=(X_1,\ldots,X_N)$ and $\vec{Y}=(Y_1,\ldots,Y_N)$ as
\begin{equation}
\frac{d}{dt}\left(\begin{array}{c}\vec{X}\\\vec{Y}\end{array}\right)=\mathbf{S}\left(\begin{array}{c}\vec{X}\\\vec{Y}\end{array}\right)+U\,\mathbf{Z} \,\, ,
\label{eq:nonlinearmodelisingmachine01bis008}
\end{equation}
with linear evolution matrix $\mathbf{S}$ and nonlinear term $\mathbf{Z}$
\begin{equation}
\mathbf{S}=\left(\begin{array}{cc}-\frac{\gamma}{2}\mathbbb{1} & -G\mathbbb{1}-\mathbf{K}\\-G\mathbbb{1}+\mathbf{K}&-\frac{\gamma}{2}\mathbbb{1}\end{array}\right) ,  \qquad \mathbf{Z}=\left(\begin{array}{c}\vec{Z}_Y \\-\vec{Z}_X\end{array}\right) \,.
\label{eq:nonlinearmodelisingmachine01bis009}
\end{equation}
In Eq.~\eqref{eq:nonlinearmodelisingmachine01bis009}, the $j$th component of the vectors $\vec{Z}_{X}$ and $\vec{Z}_{Y}$ are ${|A_j|}^2X_j$ and ${|A_j|}^2Y_j$ respectively, while we have defined $\mathbf{K}=\Delta\mathbbb{1}+\mathbf{J}/2$, with $\mathbbb{1}$ being the identity matrix.

The phases $\varphi_j$ of the KPOs can, in general, take any value in the interval $(-\pi,\pi]$. Accordingly, one can associate with the KPO network a classical energy which, in the case of homogeneous amplitudes ($R_j=R_k$ for all $j$ and $k$), is proportional to the energy of planar (XY) spins,
$E_{\rm XY}(\varphi_1,\ldots,\varphi_N)=-\sum_{j,k=1}^{N}J_{jk}\,\cos\left(\varphi_j-\varphi_k\right)$~\cite{goto2019kpoandopo}. 
However, owing to phase binarization in the steady state [Eq.~\eqref{eq:nonlinearmodelisingmachine01bis006}], the system reaches a fixed point where $\varphi_j-\varphi_k\simeq 0,\pm\pi$. In this regime, one can define effective Ising spins as $\sigma_j={\rm sign}(\varphi_j)$. With this identification, the term $E_{\rm XY}$ reduces to the classical Ising energy.
\begin{equation}
E_{\rm Ising}(\vec{\sigma})=-\sum_{j,k=1}^{N}J_{jk}\,\sigma_j\,\sigma_k \,\,. 
\label{eq:singlekerrparametrioscillatorising1}
\end{equation}

\textit{Mean-field analysis and state selection} -
We now analyze the behavior of the system close to threshold, in the linear regime, where the nonlinear term $\mathbf{Z}$ can be neglected ($U=0$). In this case, the dynamics in Eq.~\eqref{eq:nonlinearmodelisingmachine01bis008} is governed solely by $\mathbf{S}$.  
The $2N$ eigenvalues $\{\lambda^{(\pm)}_q\}$ of the matrix $\mathbf{S}$ are found from the characteristic polynomial using the partitioning of the determinant~\cite{zwillinger2002crc}:
\begin{equation}
\lambda^{(\pm)}_q=-\frac{\gamma}{2}\pm\sqrt{G^2-z_q} \,\, ,
\label{eq:nonlinearmodelisingmachine04bis04}
\end{equation}
where $\{z_q\}$ are the eigenvalues of $\mathbf{K}^2$, which are related to the eigenvalues $\{c_q\}$ of $\mathbf{J}$ by $z_q={(\Delta+c_q/2)}^2$. We define the oscillation threshold $G_{\rm th}$ as the smallest $G$ for which the real part of the largest eigenvalue turns from negative to positive.

From Eq.~\eqref{eq:nonlinearmodelisingmachine04bis04}, the largest eigenvalue is given by $\lambda_{\rm max}=-\gamma/2+\sqrt{G^2-z_{\rm min}}$, where $z_{\rm min}={\rm min}_q\{z_q\}$. The threshold condition $\lambda_{\rm max}=0$ gives the relation between $G_{\rm th}$ and $z_q$ as
\begin{equation}
G_{\rm th}=\sqrt{\frac{\gamma^2}{4}+z_{\rm min}} \,\, .
\label{eq:nonlinearmodelisingmachine04bis05}
\end{equation}
Notice that the eigenvectors $\mathbf{V}$ of $\mathbf{J}$ are also eigenvectors of $\mathbf{K}$ (and consequently of $\mathbf{K}^2$). The determination of the Ising state to which the KPO network converges at threshold is found as follows. When phase binarization $\varphi_j-\varphi_k\simeq0,\pm\pi$ is achieved, the separation in Eq.~\eqref{eq:nonlinearmodelisingmachine01bis008} of the dynamics of phases and amplitudes of $A_j=R_je^{i\varphi_j}$ yields
\begin{subequations}
\begin{align}
\frac{dR_j}{dt}&\simeq-GR_j\sin(2\varphi_j)-\frac{\gamma}{2}R_j\\
\frac{d\varphi_j}{dt}&\simeq\Delta-G\cos(2\varphi_j)+\frac{1}{2}\sum_{k=1}^{N}J_{jk}\frac{R_k}{R_j}\sigma_j\sigma_k \,\, .
\end{align}
\label{eq:nonlinearmodelisingmachinekerroscillators07}
\end{subequations}
The stationary state for $R_j$ with $R_j\neq0$, found by imposing $dR_j/dt=0$, gives the fixed-point value of the phases, which is such that $\sin(2\varphi_j)=-\gamma/2G$, for all $j$. Furthermore the condition $d\varphi_j/dt=0$, by defining the vector $\vec{R}=(R_1,\ldots,R_N)$ and the matrix $\mathbf{\Sigma}={\rm diag}(\sigma_1,\ldots,\sigma_N)$, yields the equation
\begin{equation}
\left(\Delta\mathbbb{1}+\frac{\mathbf{J}}{2}\right)\mathbf{\Sigma}\vec{R}\equiv\mathbf{K}\,\mathbf{\Sigma}\vec{R}=\left(\pm\sqrt{G^2-\frac{\gamma^2}{4}}\,\right)\mathbf{\Sigma}\vec{R} \,\,,
\label{eq:nonlinearmodelisingmachinekerroscillators021}
\end{equation}
which, for $G=G_{\rm th}$ in Eq.~\eqref{eq:nonlinearmodelisingmachine04bis05}, reads $\mathbf{K}\,\mathbf{\Sigma}\vec{R}=\pm\sqrt{z_{\rm min}}\,\mathbf{\Sigma}\vec{R}$, i.e., $\mathbf{K}^2\,\mathbf{\Sigma}\vec{R}=z_{\rm min}\,\mathbf{\Sigma}\vec{R}$. Therefore, at threshold, $\mathbf{\Sigma}\vec{R}$ is eigenvector of $\mathbf{K}^2$ with eigenvalue $z_{\rm min}$. Since $R_j>0$ for all $j$, one has that ${\rm sign}(\sigma_jR_j)\equiv\sigma_j$, encodes the Ising state at threshold $\vec{\sigma}_{\rm th}$. We stress that this phenomenology differs from the one found in degenerate OPOs at the parametric resonance ($\Delta = 0$), where the Ising state at threshold can be defined from the sign of the eigenvectors $\vec{v}_{\rm max}$, or linear combinations thereof, of the coupling matrix $\mathbf{J}$ associated to the largest eigenvalue~\cite{PhysRevE.95.022118,PhysRevLett.126.143901}. In the case of KPO in Eq.~\eqref{eq:nonlinearmodelisingmachinekerroscillators021}, the system selects the eigenvectors of $\mathbf{J}$ associated to the smallest eigenvalue of $\mathbf{K}^2$, which depends on both $\mathbf{J}$ \emph{and} $\Delta$. This key property is what gives the remarkable possibility to tune $\Delta$ to select the Ising state at threshold (Fig.~\ref{fig:networkandisingstateselection01}\textbf{b}).

\begin{figure}[t]
\centering
\includegraphics[width=8.6cm]{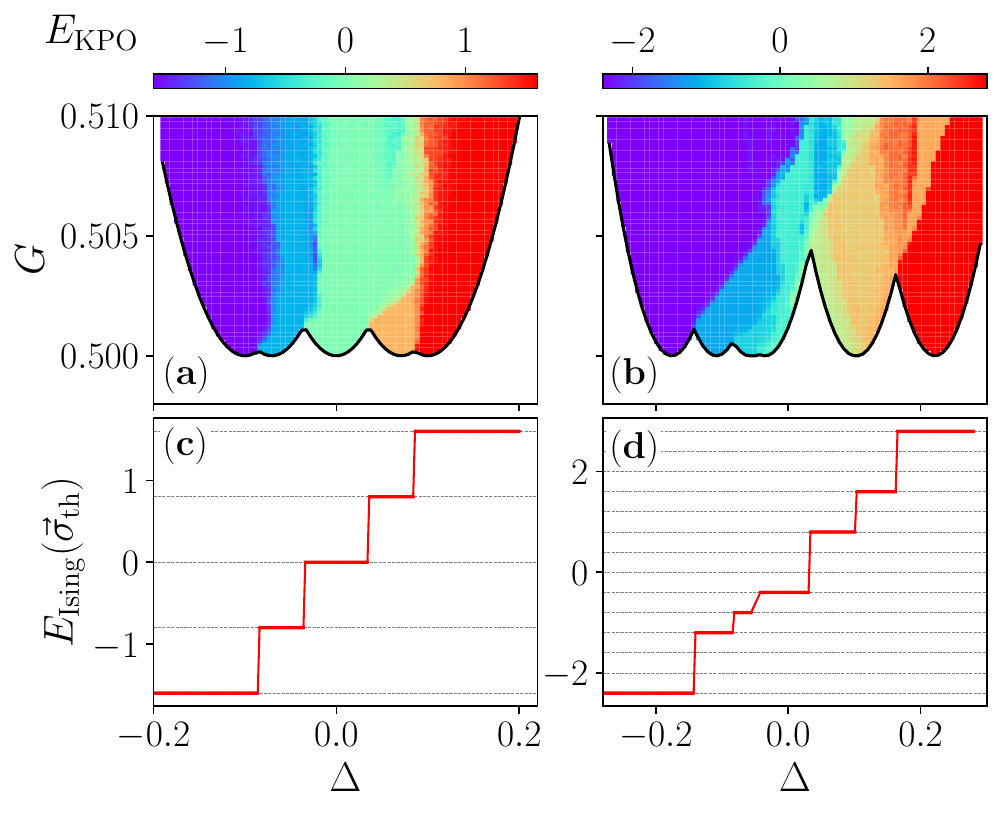}
\caption{Mean-field analysis of the KPO energy. \textbf{(a)},\textbf{(b)} Colormap of the average Ising energy $E_{\rm KPO}$ in the $G$ vs. $\Delta$ plane (color coding as in the colorbar), obtained by repeating the dynamics $50$ times with different initial conditions from Eq.~\eqref{eq:singlekerrparametrioscillatorising1}, with $\sigma_j={\rm sign}(\varphi_j)$. The KPO phases $\varphi_j$ are extracted from the steady state of Eqs.~\eqref{eq:nonlinearmodelisingmachine01bis008}. Numerical parameters: $N=8$, $U=10^{-2}$, $\gamma=1$, and $\mathbf{J}$ corresponding to a ferromagnetic chain in panel \textbf{(a)} and to a random binary sparse graph in panel \textbf{(b)} (see text). The black solid line indicates the oscillation threshold $G_{\rm th}$ [Eq.~\eqref{eq:nonlinearmodelisingmachine04bis05}]. Data for $G<G_{\rm th}$, where all oscillators relax to zero amplitude, are omitted from the plot (white region). \textbf{(c)},\textbf{(d)} Ising energy of the state selected at threshold as a function of $\Delta$, obtained from the eigenvector of $\mathbf{J}$ associated with the minimum eigenvalue of $\mathbf{K}^2$ [red data, see Eq.~\eqref{eq:nonlinearmodelisingmachinekerroscillators021} and related text], for the same $\mathbf{J}$ as in the corresponding top panel. Horizontal black dashed lines indicate the Ising energies of the full Ising spectrum. For the ferromagnetic chain, all Ising states can be selected from the eigenvectors of $\mathbf{J}$, whereas for the random graph some energies are not accessible.}
\label{fig:meanfieldanalysis01}
\end{figure}

\begin{figure}[t]
\centering
\includegraphics[width=8.4cm]{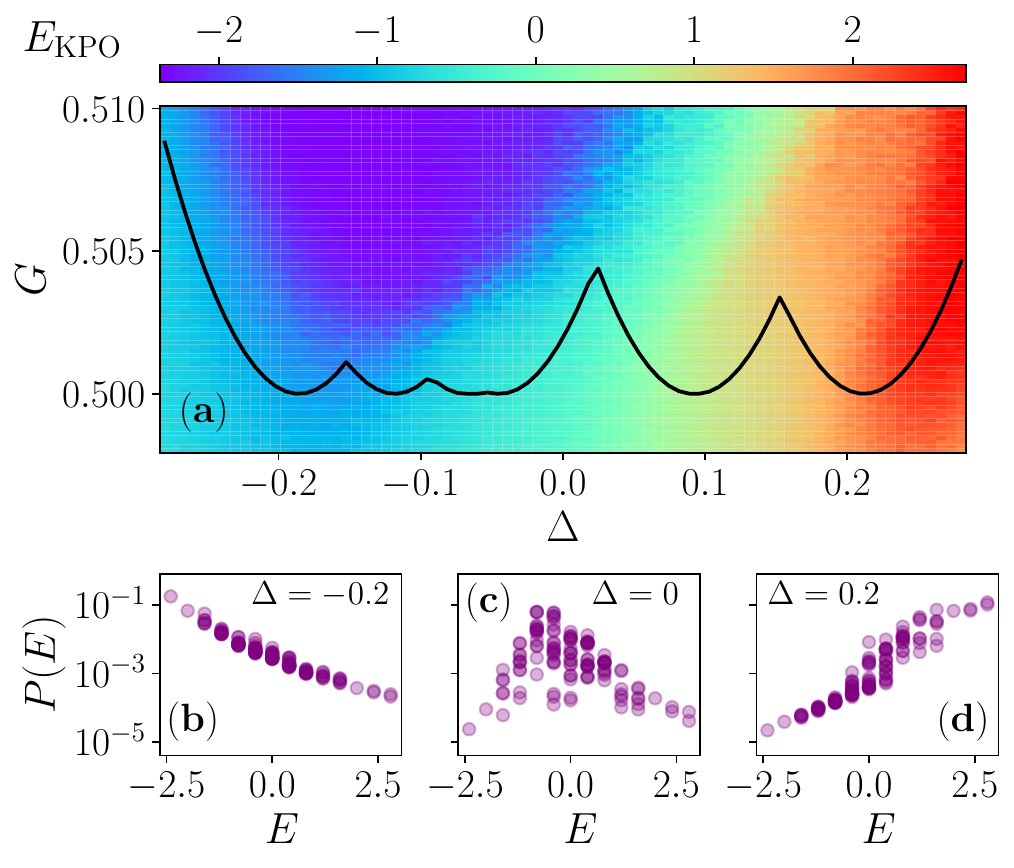}
\caption{Statistical analysis of the average energy $E_{\rm KPO}$ and probability distribution function $P(E)$ for the random graph analyzed in Fig.~\ref{fig:meanfieldanalysis01}\textbf{(b)},\textbf{(d)}. The energy is computed from Eq.~\eqref{eq:singlekerrparametrioscillatorising1} with Ising spins $\sigma_j={\rm sign}[{\rm arg}(\alpha_j)]$, where $\alpha_j$ are determined from the TWA simulations [Eq.~\eqref{eq:truncatedwignerapproximation01}]. \textbf{(a)} Colormap of $E_{\rm KPO}$ in the $G$ vs. $\Delta$ plane, and \textbf{(b)}-\textbf{(d)} $P(E)$ in log-linear scale for $\Delta=-0.2,0,0.2$, respectively, as in the panel labels, and $G$ set $0.1\%$ above $G_{\rm th}$. In agreement with Fig.~\ref{fig:meanfieldanalysis01}, the maximum of $P(E)$ is found at $E$ corresponding to the minimum and maximum Ising energy level for $\Delta=-0.2$ and $\Delta=0.2$, respectively, while for $\Delta=0$, $P(E)$ is peaked around $E$ corresponding to an intermediate Ising energy level.}
\label{fig:noisysimulationprobabilitydistribution01}
\end{figure}

The Ising state defined from the sign of $\vec{v}_{\rm max}$ yields an Ising configuration whose energy is, in general, close to the actual minimum of Eq.~\eqref{eq:singlekerrparametrioscillatorising1} (or exactly equal to the minimum in some easy cases)~\cite{yamamoto2020isingmachine,PhysRevLett.126.143901}. 
From the relation between $z_q$ and $c_q$ one sees that the KPO network at threshold converges to $\vec{v}_{\rm max}$ for sufficiently large and negative $\Delta$, thereby selecting an Ising state that is close to the \emph{minimum} of $E_{\rm Ising}$ in Eq.~\eqref{eq:singlekerrparametrioscillatorising1}. Conversely, a sufficiently large and positive $\Delta$ steers the near-threshold dynamics towards the eigenvector of $\mathbf{J}$ associated to the smallest eigenvalue, whose sign gives an Ising state with energy close to the \emph{maximum} of $E_{\rm Ising}$. For intermediate values of $\Delta$, the KPO network at threshold converges to some intermediate energy states of $E_{\rm Ising}$.

The numerical results from the mean-field analysis are shown in Fig.~\ref{fig:meanfieldanalysis01}. Panels \textbf{(a)} and \textbf{(b)} show the steady-state KPO energy $E_{\rm KPO}$ in the $G$ vs. $\Delta$ plane, computed from Eq.~\eqref{eq:singlekerrparametrioscillatorising1} with $\sigma_j={\rm sign}(\varphi_j)$. Data are reported only above the oscillation threshold [Eq.~\eqref{eq:nonlinearmodelisingmachine04bis05}, black solid line] since below threshold all KPOs converge to zero and no Ising state is defined. The phases are obtained by integrating Eq.~\eqref{eq:nonlinearmodelisingmachine01bis008} starting from random initial conditions for $X_j(0)$ and $Y_j(0)$, with $N=8$, $U=10^{-2}$, and $\gamma=1$.

We consider two prototype graphs. In panel \textbf{(a)}, the coupling matrix $\mathbf{J}$ describes a ferromagnetic periodic chain with $J_{j,k}=J$ for $|j-k|=1$ and $J=0.1$ and zero otherwise, whose Ising spectrum can be determined analytically.
Panel \textbf{(b)} encodes instead a random graph with $J_{jk}=0,\pm J$ with edge density $0.8$, and sign of the nonzero elements chosen with equal probability~\cite{PhysRevLett.132.017301}. The data display clearly distinguished regions with energy that increases with $\Delta$ as in the colorbar. Since close to threshold the system dynamics deterministically converges to suitable combinations of the eigenvectors of $\mathbf{J}$~\cite{PhysRevLett.126.143901}, the transition between region at different energy is rather sharp. As $G$ increases above threshold, the system enters the nonlinear regime. The dynamics probabilistically explores different states, and then the transition between different energy regions becomes smoother.

A deeper insight is provided by panels \textbf{(c)} and \textbf{(d)}, which show the energy in Eq.~\eqref{eq:singlekerrparametrioscillatorising1} computed from the Ising state at threshold $\vec{\sigma}_{\rm th}$ [linear regime, see Eq.~\eqref{eq:nonlinearmodelisingmachinekerroscillators021}] as a function of $\Delta$, for the same $\mathbf{J}$ as in the corresponding top panel. For each graph, all Ising energies from Eq.~\eqref{eq:singlekerrparametrioscillatorising1} are overlapped as black dashed horizontal lines. In both cases, the Ising energy at threshold increases throughout the Ising spectrum with increasing $\Delta$, thereby demonstrating state selection at threshold. For the ferromagnetic chain, the eigenvectors of $\mathbf{J}$ are sufficient to select all Ising levels already in the linear regime. Instead, for the random graph, the state selection at threshold allows to cover only a part of the Ising states.

\textit{Stochastic regime} -
Next, we study the dynamics beyond the mean-field approximation. Since we are in regime with low non-linearity ($U/\gamma \ll 1$), we employ the TWA. This semi-classical method, arising from the truncation of the third order terms in the Fokker-Plank equation associated to the system dynamics, allows us to study the KPOs via a system of coupled stochastic differential equations.  
By defining $\alpha_j\equiv\langle\hat a_j\rangle$, Eq.~\eqref{eq:nonlinearmodelisingmachine01bis007} gets modified to 
\begin{equation}
\begin{aligned}
\frac{d \alpha_j}{dt} &=  i [\Delta - U (|\alpha_j|^2 -1)] - \frac{\gamma}{2}\alpha_j -i \alpha_j^* G\\
& +\frac{i}{2} \sum_{k =1}^N J_{jk} \alpha_k + \sqrt{\frac{\gamma}{2}} \ \chi(t) \,\, ,
\end{aligned}
\label{eq:truncatedwignerapproximation01}
\end{equation}
where $\chi (t)$ is a complex Gaussian white noise such that $\braket{\chi(t) \chi( t')} = 0$ and $\braket{\chi(t) \chi^*( t')} = \delta(t-t')$.  
Compared to a classical approach with injected thermal noise~\cite{vandersande2022}, this method naturally emulates quantum fluctuations with self-regulated amplitude while keeping the numerical complexity tractable for the simulated values of $N$.

Figure~\ref{fig:noisysimulationprobabilitydistribution01} shows the numerical results obtained from the TWA analysis for the same system of $N=8$ KPO in panel \textbf{(b)} of Fig.~\ref{fig:meanfieldanalysis01}. As expected, the presence of noise does not allow to resolve the sharp transitions found instead in the mean-field case, however the energetic macro-structure of the landscape is preserved. This fact is further corroborated by the numerical results of the probability distribution function $P(E)$, which are shown in panels \textbf{(b)}, \textbf{(c)} and \textbf{(d)} for three different values of $\Delta$ and $G$ set $0.1\%$ above the mean-field oscillation threshold $G_{\rm th}$ [black line in panel \textbf{(a)}]. We numerically obtain $P(E)$ by sorting the occurrences of the KPO phase configurations during the stochastic dynamics that correspond to a given value of Ising energy $E$ in Eq.~\eqref{eq:singlekerrparametrioscillatorising1} with $\sigma_j={\rm sign}[\rm arg(\alpha_j)]$. To do so, we repeat ten times the simulation of Eq.~\eqref{eq:truncatedwignerapproximation01} up to a final time $10,000\,\gamma^{-1}$, sampling every $t =\gamma^{-1}$, for a total of $100,000$ sampled points. Notice that this high number of sampling is needed to reliably reconstruct $P(E)$.

For low values of detuning $\Delta$ (Fig.~\ref{fig:noisysimulationprobabilitydistribution01}\textbf{b}, which is for $\Delta=-0.2$), $P(E)$ resembles a Boltzmann distribution (note the log-linear scale). In particular, the Ising ground state emerges with the largest probability. A similar phenomenology was also reported for OPOs subject to Gaussian noise~\cite{Tosca2025}. However, this behavior breaks down for higher values of $\Delta$, where the most probable configuration is shifted at values of $E$ corresponding to excited states among the energetic spectrum. To well illustrate this fact, we show in panel \textbf{(c)} the result for $\Delta = 0$, where $P(E)$ is peaked at $E$ around the middle of the Ising spectrum. By further increasing the detuning to $\Delta=0.2$ [panel \textbf{(d)}], we recover a $P(E)$ that exponentially favors the highest excited state. Here, we perform noisy simulations via the TWA, thus fixing via the stochastic Langevin mapping of the Fokker-Plank equation the magnitude of the noise. We remark that an equivalent effect can be obtained by applying a classical noise to the mean-field dynamics, with properly tuned amplitude. 

This confirms the use the KPO network as an Ising selector machine even in the presence of noise. We remark that the KPOs may find energies that are slightly above the minimum or slightly below the maximum of the Ising spectrum. This is well known in graphs of increasing complexity because the near-threshold regime is insufficient to fully encode the Ising solutions for graphs of increasing complexity~\cite{PhysRevLett.126.143901}.

\textit{Conclusions} -
We demonstrated that a network of dissipatively-driven KPOs works as an Ising selector machine. We showed that varying the pump-cavity detuning $\Delta$ near the oscillation threshold steers the dynamics toward a phase configuration whose sign encodes an Ising state at a specific energy level of the spectrum defined by the coupling matrix $\mathbf{J}$. For low and high detuning the system converges close to the Ising lowest or highest energy, while for intermediate values of the detuning the system selects Ising states in between the spectrum. The allowed states that can be selected depends on the eigenvectors of the linear evolution matrix.

We confirmed our results beyond the mean-field level, by simulating the KPO dynamics in the presence of quantum noise. By resorting to the truncated Wigner approximation method, we showed that even in the presence of noise the energetic structure of the landscape is preserved. The targeted state emerges with an exponentially enhanced probability over the rest of the Ising spectrum. An interesting perspective is the study of the fully quantum regime using non-semiclassical techniques as, for example, the variational multi-Gaussian~\cite{tosca2025efficientvariationaldynamicsopen, tosca2026variationaldynamicsopenquantum}.

In this work we focus on KPOs for concreteness and for their direct experimental relevance. We remark that the reported state-selection principle is rooted in the interplay between the detuning and the spectral properties of the coupling matrix, a mechanism that could be explored in other driven-dissipative platforms hosting parametric oscillators. Our work extends the paradigm of oscillator networks as Ising simulators, allowing to controllably select the Ising state, much beyond the conventional idea of Ising machine for combinatorial optimization.

An intriguing perspective is the realization of engineered state-selection methods in high-dimensional vector spin models with tunable dimension, such as hyperspins~\cite{strinati2022hyperspinmachine} with OPOs, laying the foundation for using classical and quantum oscillator networks to tackle general, high-dimensional optimization problems.

\textit{Acknowledgements} - Funded by the European Union (HORIZON-ERC-2023-ADG HYPERSPIM project Grant No. 101139828).

\end{document}